\documentclass[runningheads]{llncs}
\usepackage{amsfonts}

\usepackage{latexsym}
\usepackage{amssymb}
\usepackage{epsfig}
\usepackage{amsmath}
\usepackage[mathscr]{eucal}
\usepackage{subfigure}
\usepackage{graphicx}

\newcommand{\ind}{1\hspace{-2.3mm}{1}}

\begin{document}

\title{Achieving positive rates with predetermined dictionaries}
\titlerunning{Achieving positive rates}

\author{ \textbf{Ghurumuruhan Ganesan}}
\authorrunning{G. Ganesan}
\institute{Institute of Mathematical Sciences, HBNI, Chennai\\
\email{gganesan82@gmail.com }}

\date{}
\maketitle

\begin{abstract}
In the first part of the paper we consider binary input channels that are not necessarily stationary and show how positive rates can be achieved using codes constrained to be within predetermined dictionaries. We use a Gilbert-Varshamov-like argument to obtain the desired rate achieving codes. Next we study the corresponding problem for channels with arbitrary alphabets and use conflict-set decoding to show that if the dictionaries are contained within ``nice" sets, then positive rates are achievable.

\vspace{0.1in} \noindent \textbf{Key words:} Positive rates, predetermined dictionaries.

\vspace{0.1in} \noindent \textbf{AMS 2000 Subject Classification:} Primary:
94A15, 94A24.
\end{abstract}

\bigskip

\setcounter{equation}{0}
\renewcommand\theequation{\arabic{section}.\arabic{equation}}
\section{Introduction} \label{intro}
Achieving positive rates with low probability of error in communication channels is an important problem in information theory~\cite{cov}. In general, a rate~\(R\) is defined to be achievable if there exists codes with rate~\(R\) and having arbitrarily small error probability as the code length~\(n \rightarrow \infty.\) The existence of such codes is determined through the probabilistic method of choosing a random code (from the set of all possible codes) and showing that the chosen code has small error probability.

In many cases of interest, we would like to select codes satisfying certain constraints or equivalently from a predetermined \emph{dictionary} (see~\cite{bernd,ram} for examples). For stationary channels, the method of types~\cite{csi,csi2} can be used to study positive rate achievability with the restriction that the dictionary falls within the set of words belonging to a particular type.  In this paper, we study achievability of positive rates with \emph{arbitrary} deterministic dictionaries for both binary and general input channels using counting techniques.


The paper is organized as follows: In Section~\ref{sec_bin}, we study positive rate achievability in binary input channels using predetermined dictionaries. Next in Section~\ref{pre_det}, we describe the rate achievability problem for arbitrary stationary channels and state our result Theorem~\ref{thm3} regarding achieving positive rates using given dictionaries. Finally, in Section~\ref{pf_thm}, we prove Theorem~\ref{thm3}.


\setcounter{equation}{0}
\renewcommand\theequation{\arabic{section}.\arabic{equation}}
\section{Binary Channels}\label{sec_bin}
For integer~\(n \geq 1,\) an element of the set~\(\{0,1\}^{n}\) is said to be a  \emph{codeword} or simply word, of length~\(n.\)
Consider a discrete memoryless symmetric channel with input alphabet~\(\{0,1\}^{n}\) that corrupts a transmitted word~\(\mathbf{x} = (x_1,\ldots,x_n)\) as follows. If~\(\mathbf{Y} = (Y_1,\ldots,Y_n)\) is the received (random) word, then
\begin{equation}\label{rand_noise}
Y_i := x_i \ind(W_i=0) + (1-x_i) \ind(W_i = 1) + \varepsilon \ind(W_i = \varepsilon)
\end{equation}
for all~\(1 \leq i \leq n,\) where~\(\ind(.)\) denotes the indicator function and~\(\varepsilon\) denotes the erasure symbol. If~\(W_i = 1,\)
then the bit~\(x_i\) is substituted and if~\(W_i = \varepsilon,\) then~\(x_i\) is erased. The random variables~\(\{W_i\}_{1 \leq i \leq n}\) are independent with
\begin{equation}\label{w_p_i}
\mathbb{P}(W_i =1) = p_f(i) \text{ and } \mathbb{P}(W_i = \varepsilon)=p_e(i)
\end{equation}
and so the probability of a bit error (due to either a substituted bit or an erased bit) at ``time" index~\(i\) is~\(p_f(i) + p_e(i).\)
Letting~\(\mathbf{W} := (W_1,\ldots,W_n),\) we also denote~\(\mathbf{Y} =: h(\mathbf{x},\mathbf{W})\) where~\(h\) is a deterministic function defined via~(\ref{rand_noise}). We are interested in communicating through the above described channel, with low probability of error, using words from a predetermined (deterministic) dictionary.

\subsection*{Dictionaries}
A dictionary of size~\(M\) is a set~\({\cal D} \subseteq \{0,1\}^{n}\) of cardinality~\(\#{\cal D} = M.\) A subset~\({\cal C}  = \{\mathbf{x}_1,\ldots,\mathbf{x}_L\}\subseteq {\cal D}\) is said to be an~\(n-\)length \emph{code} of size~\(L,\) contained in the dictionary~\({\cal D}.\) Suppose we transmit a word picked from~\({\cal C},\) through the channel given by~(\ref{rand_noise}) and receive the (random) word~\(\mathbf{Y}.\) Given~\(\mathbf{Y}\) we would like an estimate~\(\hat{\mathbf{x}}\) of the word from~\({\cal C}\) that was transmitted.
A decoder~\(g : \{0,1,\varepsilon\}^{n} \rightarrow {\cal C}\) is a deterministic map that uses the received word~\(\mathbf{Y}\) to obtain an estimate of the transmitted word. The probability of error corresponding to the code~\({\cal C}\) and the decoder~\(g\) is then defined as
\begin{equation}\label{dec_err2}
q({\cal C},g) := \max_{1 \leq i \leq L} \mathbb{P}\left(g\left(h\left(\mathbf{x}_i,\mathbf{W}\right)\right) \neq \mathbf{x}_i\right),
\end{equation}
where~\(\mathbf{W} = (W_1,\ldots,W_n)\) is the additive noise as described in~(\ref{rand_noise}).

We have the following definition regarding achievable rates using predetermined dictionaries.
\begin{definition}
Let~\(R > 0\) and let~\({\cal F} := \{{\cal D}_n\}_{n \geq 1}\) be any sequence of dictionaries such that each~\({\cal D}_n\) has size at least~\(2^{nR}.\) We say that~\(R > 0\) is an~\({\cal F}-\)\emph{achievable} rate if the following holds true for every~\(\epsilon  >0:\) For all~\(n\) large, there exists a code~\({\cal C}_n \subset {\cal D}_n\) of size~\(\#{\cal C}_n = 2^{nR}\) and a  decoder~\(g_n\) such that the probability of error~\(q({\cal C}_n,g_n) < \epsilon.\)
\end{definition}
If~\({\cal D}_n = \{0,1\}^{n}\) for each~\(n,\) then the above reduces to the usual concept of rate achievability as in~\cite{cov} and we simply say that~\(R\) is achievable. 


For~\(0 < x < 1\) we define the entropy function
\begin{equation}\label{hp_def}
H(x) := -x\cdot \log{x} - (1-x) \cdot \log(1-x),
\end{equation}
where all logarithms in this section to the base two and have the following result.
\begin{theorem}\label{main_thm2} For integer~\(n \geq 1\) let~\[\mu_f  = \mu_f(n) := \sum_{i=1}^{n}p_f(i) \text{ and }\mu_e = \mu_e(n) := \sum_{i=1}^{n} p_e(i)\] be the expected number of bit substitutions and erasures, respectively in an~\(n-\)length codeword and suppose
\begin{equation}\label{mu_n_cond}
\min\left(\mu_f(n), \mu_e(n)\right) \longrightarrow \infty \text{ and } p := \limsup_n \frac{1}{n}\left(2\mu_f(n) + \mu_e(n)\right) <\frac{1}{2}.
\end{equation}
Let~\(H(p) < \alpha \leq 1\) and let~\({\cal F} := \{{\cal D}_n\}_{n \geq 1}\) be any sequence of dictionaries satisfying~\(\#{\cal D}_n \geq 2^{\alpha n},\) for each~\(n.\) We have that every~\(R < \alpha - H(p)\) is~\({\cal F}-\)achievable.
\end{theorem}
For a given~\(\alpha,\) let~\(p(\alpha)\) be the largest value of~\(p\) such that~\(H(p) <\alpha.\) The above result says that every~\(R < \alpha-H(p)\) is achievable using arbitrary dictionaries. We use Gilbert-Varshamov-like arguments to prove Theorem~\ref{main_thm2} below.


As a special case, for binary symmetric channels with crossover probability~\(p_f,\) each bit is independently substituted with probability~\(p_f.\) No erasures occur and so~\[\mu_f(n) = np_f \text{ and } \mu_e(n) = 0.\] Thus~\(p = 2p_f\) and from Theorem~\ref{main_thm2} we therefore have that if~\(H(2p_f) < \alpha,\) then every~\(R < \alpha-H(2p_f)\) is achievable.

\subsection*{Proof of Theorem~\ref{main_thm2}}
The main idea of the proof is as follows. Using standard deviation estimates, we first obtain an upper bound on the number
of possible errors that could occur in a transmitted word. More specifically, if~\(T\) denotes the number of bit errors
in an~\(n-\)length word and~\(\epsilon >0\) is given, we use standard deviation estimates to determine~\(T_0 = T_0(n)\) such that~\(\mathbb{P}(T > T_0) \leq \epsilon.\) We then use a Gilbert-Varshamov argument to obtain a code that can correct up to~\(T_0\) bit errors.
The details are described below.

We prove the Theorem in two steps. In the first step, we construct the code~\({\cal C}\) and decoder~\(g\)
and in the second step, we estimate the probability of the decoding error for~\({\cal C}\) using~\(g.\)
For~\(\mathbf{x},\mathbf{y} \in \{0,1\}^{n},\) we let~\(d_H(\mathbf{x},\mathbf{y}) = \sum_{i=1}^{n} \ind(x_i \neq y_i)\)
be the Hamming distance between~\(\mathbf{x}\) and~\(\mathbf{y},\) where as before~\(\ind(.)\) denotes the indicator function.
The minimum distance of a code is the minimum distance between any two words in a code.


\emph{\underline{Step 1}}: Assume for simplicity that~\(t := np(1+2\epsilon)\) is an integer and let~\(d = t+1.\) For a word~\(\mathbf{x}\) let~\(B_{d-1}(\mathbf{x})\) be the set of words that are at a distance of at most~\(d-1\) from~\(\mathbf{x}.\) If~\({\cal C} \subseteq {\cal D}\) is a maximum size code with minimum distance at least~\(d,\) then by the maximality of~\({\cal C}\) we must have
\begin{equation}\label{gv_arg}
\bigcup_{\mathbf{x} \in {\cal C}} B_{d-1}(\mathbf{x}) = {\cal D}.
\end{equation}
This is known as the Gilbert-Varshamov argument~\cite{huff}.

The cardinalities of~\({\cal D}\) and~\(B_{d-1}(\mathbf{x})\) are~\(2^{\alpha n}\) and~\(\sum_{i=0}^{d-1} {n \choose i}\) respectively and so from~(\ref{gv_arg}), we see that the code~\({\cal C} \) has size
\begin{equation}\label{gv_bound}
\#{\cal C} \geq \frac{2^{\alpha n}}{\sum_{i=0}^{d-1} {n \choose i}}
\end{equation}
and minimum distance at least~\(d.\) Also since~\(p < \frac{1}{2},\) we have for all small~\(\epsilon  >0\) that~\({n \choose i} \leq {n \choose d-1} = {n \choose np(1+2\epsilon)}\) and so~\(\sum_{i=0}^{d-1} {n \choose i} \leq n \cdot {n \choose np(1+2\epsilon)}.\) Using Stirling approximation we get~\[{n \choose np(1+2\epsilon)} \leq 4en\cdot 2^{nH(p+2p\epsilon)}\] and so from~(\ref{gv_bound}), we get for~\(\delta > 0\) that
\begin{equation}\label{gv_bound2}
\#{\cal C} \geq \frac{1}{4en^2} \cdot 2^{n(\alpha-H(p+2p\epsilon))} \geq 2^{n(\alpha-H(p)-\delta)}
\end{equation}
provided~\(\epsilon  >0\) is small.

We now use a two stage decoder described as follows: Suppose the received word is~\(\mathbf{Y}\)
and for simplicity suppose that the last~\(e\) positions in~\(\mathbf{Y}\) have been erased.
For a codeword~\(\mathbf{x} = (x_1,\ldots,x_n),\) let~\(\mathbf{x}_{red} := (x_1,\ldots,x_{n-e})\)
be the reduced word formed by the first~\(n-e\) bits. Let~\({\cal C}_{red} = \{\mathbf{x}_{red} : \mathbf{x} \in {\cal C}\}\) be the set of all reduced codewords in the code~\({\cal C}\) formed by the first~\(n-e\) bits.

In the first stage of the decoding process, the decoder corrects bit substitutions
by collecting all words~\({\cal S} \subseteq {\cal C}_{red}\) whose Hamming distance from~\(\mathbf{Y}_{red}\) is minimum. If~\({\cal S} \) contains exactly one word, say~\(\mathbf{z}_{red},\) the decoder outputs~\(\mathbf{z}_{red}\) as the estimate obtained in the first step of the iteration. Otherwise, the decoder outputs ``decoding error". In the second stage of the decoding process, the decoder uses~\(\mathbf{z}_{red}\) to correct the erasures. Formally let~\({\cal S}_{e} := \{\mathbf{x} \in {\cal C} : \mathbf{x}_{red} = \mathbf{z}_{red}\}\) be the set of all codewords whose first~\(n-e\) bits match~\(\mathbf{z}_{red}.\) If there exists exactly one word~\(\mathbf{z}\) in~\({\cal S}_{e},\) then the decoder outputs~\(\mathbf{z}\) to be the transmitted word. Else the decoder outputs ``decoding error".

\emph{\underline{Step 2}}: Suppose a word~\(\mathbf{x} \in {\cal C}\) was transmitted and the received word is~\(\mathbf{Y}.\)
Let~\(\mathbf{W} = (W_1,\ldots,W_n)\) be the random noise vector as in~(\ref{rand_noise})
and let~\[T_f := \sum_{i=1}^{n}\ind(W_i =1)\] be the number of bits that have been substituted so that~\[\mathbb{E}T_f = \sum_{i=1}^{n}p_f(i) = \mu_f(n),\] by~(\ref{w_p_i}). By standard deviation estimates (Corollary~\(A.1.14,\) pp. 312,~\cite{alon}) we have
\begin{equation}\label{azum_app_f}
\mathbb{P}\left(|T_f-\mu_f(n)| \geq \epsilon \mu_f(n)\right) \leq 2e^{-\frac{\epsilon^2}{4}\mu_f(n)} \leq \frac{\epsilon}{2}
\end{equation}
for all~\(n\) large, by the first condition of~(\ref{mu_n_cond}). Similarly
if~\(T_e = \sum_{i=1}^{n} \ind(W_i = \varepsilon)\) is the number of erased bits, then
\begin{equation}\label{azum_app_e}
\mathbb{P}\left(|T_e-\mu_e(n)| \geq \epsilon \mu_e(n)\right) \leq 2e^{-\frac{\epsilon^2}{4}\mu_e(n)} \leq \frac{\epsilon}{2}
\end{equation}
for all~\(n\) large. 

Next, using the second condition of~(\ref{mu_n_cond}) we have that~\[(2\mu_f(n) + \mu_e(n))(1+\epsilon) \leq np(1+2\epsilon) = t\] for all~\(n\) large and so from~(\ref{azum_app_f}) and~(\ref{azum_app_e}) we get that~\(\mathbb{P}\left(2T_f + T_e \geq t\right) \leq \epsilon\) for all~\(n\) large. If~\(2T_f+T_e \leq t,\) then by construction the decoder outputs~\(\mathbf{x}\) as the estimate of the transmitted word. Therefore a decoding error occurs only if~\(2T_f + T_e \geq t\) which happens with probability at most~\(\epsilon.\) Combining with~(\ref{gv_bound2}) and using the fact that~\(\delta >0\) is arbitrary, we get that every~\(R < \alpha - H(p)\) is~\({\cal F}-\)achievable.\;\;\;\;\;\;\;\;\;\;\;\;\;\;\;\;\;\;\;\;\;\;\;\;\;\;\;\;\;\;\;\;\;
\;\;\;\;\;\;\;\;\;\;\;\;\;\;\;\;\;\;\;\;\;\;\;\;\;\;\;\;\;\;\;\;\;\;\;\;\;\;\;\;\;\;\;\;\;\;\;\;\;\;\;\;\;\;\;\;\;\;\;\;~\(\qed\)

\setcounter{equation}{0}
\renewcommand\theequation{\arabic{section}.\arabic{equation}}
\section{General channels} \label{pre_det}
Consider a discrete memoryless channel with finite input alphabet~\({\cal X}\) of size~\(N := \#{\cal X},\) a finite output alphabet~\({\cal Y}\) and a transition probability~\(p_{Y|X}(y|x), x \in {\cal X}, y \in {\cal Y}.\) The term~\(p_{Y|X}(y|x)\) denotes the probability that output~\(y\) is observed given that input~\(x\) is transmitted through the channel.

For~\(n \geq 1\) we define a subset~\({\cal D}_n \subseteq {\cal X}^{n}\) to be a \emph{dictionary}. A subset~\({\cal C} = \{x_1,\ldots,x_M\} \subseteq {\cal D}_n\) is defined to be an~\(n-\)length code contained within the dictionary~\({\cal D}_n.\) Suppose we transmit the word~\(x_1\) and receive the (random) word~\(\Gamma_{x_1} \in {\cal Y}^{n}.\) Given~\(\Gamma_{x_1}\) we would like an estimate~\(\hat{x}\) of the word from~\({\cal C}\) that was transmitted. A decoder~\(g : {\cal Y}^{n} \rightarrow {\cal C}\) is a deterministic map that ``guesses" the transmitted word based on the received word~\(\Gamma_{x_1}.\) We denote the probability of error corresponding to the code~\({\cal C}\) and the decoder~\(g\) as
\begin{equation}\label{dec_err3}
q({\cal C},g) := \max_{x \in {\cal C}} \mathbb{P}\left(g(\Gamma_{x}) \neq x\right).
\end{equation}

To study positive rate achievability using arbitrary dictionaries, we have a couple of preliminary definitions. Let~\(p_X(.)\) be any probability distribution on the input alphabet~\({\cal X}\) and let~\(H(X) := -\sum_{x \in {\cal X}} p_X(x) \log{p_X(x)}\)
be the entropy of a random variable~\(X\) where the logarithm is to the base~\(N\) here.  Let~\(Y\) be a random variable having joint distribution~\(p_{XY}(x,y)\) with the random variable~\(X\) defined by~\(p_{XY}(x,y) := p_{Y|X}(y|x) \cdot p_X(x).\) Thus~\(Y\) is the random output of the channel when the input is~\(X.\) Letting~\(p_Y(y) := \sum_{x} p_{XY}(x,y)\) be the marginal of~\(Y\) we have that the joint entropy and conditional entropy~\cite{cov} are respectively given by~\[H(X,Y) = -\sum_{x,y} p_{XY}(x,y)\log{p_{XY}(x,y)}  \] and~\[H(Y|X) = -\sum_{x,y} p_{XY}(x,y) \log{p_{Y|X}(y|x)}.\]

The following result obtains positive rates achievable with predetermined dictionaries for the channel described above.
\begin{theorem}\label{thm3} Let~\(p_X,p_Y\) and~\(p_{XY}\) be as above and let~\(0  < \alpha \leq H(X).\) For every~\(\epsilon  >0\) and for all~\(n\) large, there is a deterministic set~\({\cal B}_n\) with size at least~\(N^{n(H(X)-2\epsilon)}\) and satisfying the following property: If~\({\cal D}_n\) is any subset of~\({\cal B}_n\) with cardinality~\(N^{n(\alpha-2\epsilon)}\) and
\begin{equation}\label{r_rate3}
R < \alpha - H(Y|X) - H(X|Y)  - 7\epsilon
\end{equation}
is positive, then there exists a code~\({\cal C}_n \subset {\cal D}_n\) containing~\(N^{nR}\) words and a decoder~\(g_n\) with error probability~\(q\left({\cal C}_n,g_n\right) < \epsilon.\)
\end{theorem}
Thus if the sequence of dictionaries~\({\cal F} := \{{\cal D}_n\}_{n \geq 1}\) is such that~\({\cal D}_n \subset {\cal B}_n\) for each~\(n,\) then every~\(R < \alpha-H(Y|X)-H(X|Y)\) is~\({\cal F}-\)achievable. Also, setting~\(\alpha = H(X)\) and~\({\cal D}_n = {\cal B}_n\) also gives us that every~\(R < H(X)  - H(X|Y) - H(Y|X)\) is achievable in the usual sense of~\cite{cov}, without any restrictions on the dictionaries. For context, we remark that Theorem~\ref{main_thm2} holds for \emph{arbitrary} dictionaries.

To prove Theorem~\ref{thm3}, we use typical sets~\cite{cov} together with conflict set decoding described in the next section. Before we do so, we present an example to illustrate Theorem~\ref{thm3}.

\subsection*{Example}
Consider a binary asymmetric channel with alphabet~\({\cal X} = {\cal Y} = \{0,1\}\) and transition probability
\[p(1|0) = p_0 = 1-p(0|0) \text{ and } p(0|1) = p_1 = 1-p(1|1).\]
To apply Theorem~\ref{thm3}, we assume that the input has the symmetric distribution~\(\mathbb{P}(X_i = 0) = \frac{1}{2} = \mathbb{P}(X_i = 1)\)
so that the entropy~\(H(X)\) equals its maximum value of~\(1.\) The entropy of the output~\(H(Y) = H(q)\) where~\(q = \frac{1-p_0+p_1}{2}\) and the conditional entropies equal~\[H(Y|X) = \frac{1}{2}\left(H(p_0) + H(p_1)\right) \text{ and }H(X|Y) = \frac{1}{2}\left(H(p_0) + H(p_1)\right) + 1-H(q).\]

Set~\(p_0 = p\) and~\(p_1 = p + \Delta.\) If both~\(p\) and~\(\Delta\) are small, then~\(H(q)\) is close to one and~\(H(p_0)\) and~\(H(p_1)\) are close to zero. We assume that~\(p\) and~\(\Delta\) are such that~\[\alpha_0  := H(Y|X) +H(X|Y)=H(p) + H(p+\Delta) + 1-H\left(\frac{1-\Delta}{2}\right)\]
is strictly less than one and choose~\(\alpha > \alpha_0.\) Every~\(R < \alpha-\alpha_0\) is then~\({\cal F}-\)achievable as in the statement following Theorem~\ref{thm3} and every~\(R < 1-\alpha_0\) is achievable without any dictionary restrictions, in the usual sense of~\cite{cov}.

\begin{figure}[tbp]
\centering
\includegraphics[width=3.5in, clip=true]{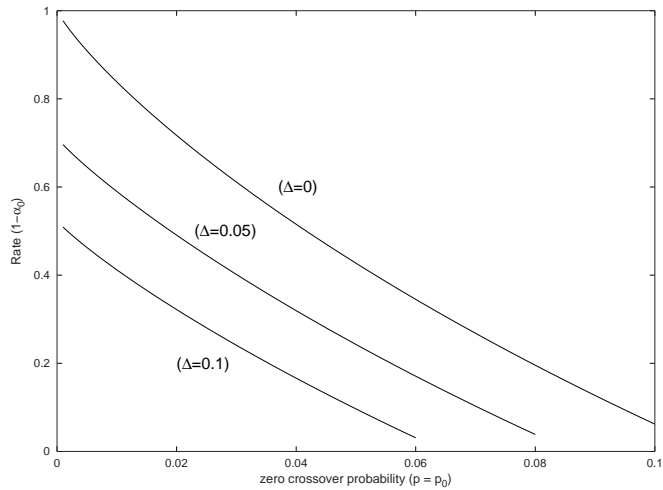}
\caption{Plotting the rate~\(1-\alpha_0\) as a function of~\(p\) for various values of the asymmetry factor~\(\Delta.\)}
\label{p_del}
\end{figure}

In Figure~\ref{p_del}, we plot~\(1-\alpha_0\) as a function of~\(p\) for various values of the asymmetry factor~\(\Delta.\) For example, for an asymmetry factor of~\(\Delta = 0.05\) we see that positive rates are achievable for~\(p\) roughly up to~\(0.08.\)

\setcounter{equation}{0}
\renewcommand\theequation{\arabic{section}.\arabic{equation}}
\section{Proof of Theorem~\ref{thm3}}\label{pf_thm}
We use conflict set decoding to prove Theorem~\ref{thm3}. Therefore in the first part of this section, we prove an auxiliary result regarding conflict set decoding that is also of independent interest.

\subsection{Conflict set decoding}
Consider a discrete memoryless channel with finite input alphabet~\({\cal X}_0\) and finite output alphabet~\({\cal Y}_0\) and transition probability~\(p_0(y|x), x \in {\cal X}_0, y \in {\cal Y}_0.\) For convenience, we define the channel by a collection of random variables~\(\theta_x, x \in {\cal X}_0\) with the distribution~\(\mathbb{P}\left(\theta_{x} = y\right) := p_0(y|x)\) for~\(y \in {\cal Y}_0.\) All random variables are defined on the probability space~\((\Omega,{\cal F}, \mathbb{P}).\) For~\(\epsilon > 0,x \in {\cal X}_0\) and~\(y \in {\cal Y}_0\) we let~\(D(x,\epsilon)\) and~\(C(y,\epsilon)\) be deterministic sets such that
\begin{equation}\label{dx_def}
\mathbb{P}\left(\theta_x \in D(x,\epsilon) \right)  \geq 1-\epsilon \text{ and } C(y,\epsilon)= \{x : y \in D(x,\epsilon)\}.
\end{equation}
We define~\(D(x,\epsilon)\) to be an~\(\epsilon-\)\emph{probable} set or simply probable output set corresponding to the input~\(x\)
and for~\(y \in {\cal Y}_0,\) we denote~\(C(y,\epsilon)\) to be the~\(\epsilon-\)\emph{conflict set} or simply conflict set corresponding to the output~\(y.\) There are many possible choices for~\(D(x,\epsilon);\) for example~\(D(x,\epsilon) = {\cal Y}_0\) is one choice.
In Proposition~\ref{main_thm} below, we show however that choosing~\(\epsilon-\)probable sets as small as possible allows us to increase the size of the desired code. We also define
\begin{equation}\label{dl_dr_def}
d_L(\epsilon) := \max_{x \in {\cal X}_0} \#D(x,\epsilon) \text{ and } d_R(\epsilon) := \max_{y \in {\cal Y}_0} \#C(y,\epsilon)
\end{equation}
where~\(\#A\) denotes the cardinality of the set~\(A.\)

As before, a \emph{code}~\({\cal C}\) of size~\(M\) is a set of distinct words~\(\{x_1,\ldots,x_M\} \subseteq {\cal X}_0.\) Suppose we transmit the word~\(x_1\) and receive the (random) word~\(\theta_{x_1}.\) Given~\(\theta_{x_1}\) we would like an estimate~\(\hat{x}\) of the word from~\({\cal C}\) that was transmitted. A decoder~\(g : {\cal Y}_0 \rightarrow {\cal C}\) is a deterministic map that guesses the transmitted word based on the received word~\(\theta_{x_1}.\) We denote the probability of error corresponding to the code~\({\cal C},\) the decoder~\(g\) and the collection of the probable sets~\({\cal D} := \{D(x,\epsilon)\}_{x \in {\cal X}_0}\) as
\begin{equation}\label{dec_err}
q({\cal C},g,{\cal D}) := \max_{x \in {\cal C}} \mathbb{P}\left(g(\theta_{x}) \neq x\right).
\end{equation}

We have the following Proposition.
\begin{proposition}\label{main_thm}  For~\(\epsilon  >0\) let~\({\cal D} =\{D(x,\epsilon)\}_{x \in {\cal X}_0}\) be any collection of~\(\epsilon-\)probable sets.  If there exists an integer~\(M\) satisfying
\begin{equation}\label{m_est}
M < \frac{\#{\cal X}_0}{d_L(\epsilon) \cdot d_R(\epsilon)},
\end{equation}
then there exists a code~\({\cal C} \subseteq {\cal X}_0\) of size~\(M\) and a decoder~\(g\) whose decoding error probability is~\(q({\cal C},g, {\cal D}) < \epsilon.\)
\end{proposition}
Thus as long as the number of words is below a certain threshold, we are guaranteed that the error probability is sufficiently small. Also, from~(\ref{m_est}) we see that it would be better to choose probable sets with as small cardinality as possible.

\subsection*{Proof of Proposition~\ref{main_thm}}
\emph{Code construction}: We recall that by definition, given input~\(x,\) the output~\(\theta_x\) belongs to the set~\(D(x,\epsilon)\) with probability at least~\(1-\epsilon.\) Therefore we first construct a code~\({\cal C} = \{x_1,\ldots,x_M\}\) containing~\(M\) distinct words and satisfying
\begin{equation}\label{disj_prob}
D(x_i,\epsilon) \cap D(x_j, \epsilon) = \emptyset \text{ for all }x_i,x_j \in {\cal C}.
\end{equation}
Throughout we assume that~\(M\) satisfies~(\ref{m_est}). To obtain the desired distinct words, we use following the bipartite graph representation. Let~\(G=G(\epsilon)\) be a bipartite graph with vertex set~\({\cal X}_0 \cup {\cal Y}_0.\)  We join~\(x \in {\cal X}_0\) and~\(y \in {\cal Y}_0\) by an edge if and only if~\(y \in D(x,\epsilon).\) The size of~\(D(x,\epsilon)\) therefore represents the degree of the vertex~\(x\) and the size of~\(C(y,\epsilon)\) represents the degree of the vertex~\(y.\) By definition (see~(\ref{dl_dr_def}))~\(d_L = \max_{x \in {\cal X}} \#D(x,\epsilon) \text{ and } d_R = \max_{y \in {\cal Y}} \#C(y,\epsilon)\) denote the maximum degree of a left vertex and a right vertex, respectively, in~\(G.\) We say that a set of vertices~\(\{x_1,\ldots,x_M\}\) is \emph{disjoint} if for all~\(i \neq j,\) the vertices~\(x_i\) and~\(x_j\) have no common neighbour (in~\({\cal Y}_0\)). Constructing codes with disjoint~\(\epsilon-\)probable sets satisfying~(\ref{disj_prob}) is therefore equivalent to finding disjoint sets of vertices in~\({\cal X}_0.\)

We now use direct counting to get a set of~\(M\) disjoint vertices~\(\{x_1,\ldots,x_M\}\) in~\({\cal X}_0.\) First we pick any vertex~\(x_1 \in {\cal X}_0.\) The degree of~\(x_1\) is at most~\(d_L\) and moreover, each vertex in~\(D(x_1,\epsilon) \subseteq {\cal Y}\) has at most~\(d_R\) neighbours in~\({\cal X}_0.\) The total number of (bad) vertices of~\({\cal X}_0\) adjacent to some vertex in~\(D(x_1,\epsilon)\) is at most~\(d_L \cdot d_R.\) Removing all these bad vertices, we are left with a bipartite subgraph~\(G_1\) of~\(G\) whose left vertex set has size at least~\(N_0 - d_L \cdot d_R\) where~\(N_0 = \#{\cal X}_0.\) We now pick one vertex in the left vertex set of~\(G_1\) and continue the above procedure. After the~\(i^{th}\) step, the number of left vertices remaining is~\(N_0 - i\cdot d_L \cdot d_R\) and so from~(\ref{m_est}) we get that this process continues at least for~\(M\) steps. The words corresponding to vertices~\(\{x_1,\ldots,x_M\}\) form our code~\({\cal C}.\) \\\\
\emph{Decoder definition}: Let~\({\cal C}\) be the code as constructed above. For decoding, we use the \emph{conflict-set decoder} defined as follows: If~\(y \in D(x_j,\epsilon)\) for some~\(x_j \in {\cal C}\) and the conflict set~\(C(y,\epsilon)\) does not contain any of word of~\({\cal C} \setminus \{x_j\},\) then we set~\(g(y) = x_j.\) Otherwise, we set~\(g(y)\) to be any arbitrary value; for concreteness, we set~\(g(y) = x_1.\)

We claim that the probability of error of the conflict-set decoder is at most~\(\epsilon.\) To see this is true, suppose we transmit the word~\(x_i.\) With probability at least\\\(1-\epsilon,\) the corresponding output~\(\theta_{x_i} \in D(x_i,\epsilon).\) Because~(\ref{disj_prob}) holds, we must necessarily have that~\(y \notin D(x_k,\epsilon)\) for any~\(k \neq j.\) This implies that the conflict-set decoder outputs the correct word~\(x_i\) with probability at least~\(1-\epsilon.\)\;\;\;\;\;\;\;\;\;\;~\(\qed\)

We now prove Theorem~\ref{thm3} using typical sets and conflict set decoding.
\subsection{Proof of Theorem~\ref{thm3}}
For notational simplicity we prove Theorem~\ref{thm3} with~\({\cal X}  = {\cal Y} = \{0,1\}.\) An analogous analysis holds for the general case.

The proof consists of three steps. In the first step, we define and estimate the occurrence of certain typical sets. In the next step, we use the typical sets constructed in Step~\(1\) to determine the set~\({\cal B}_n\) in the statement of the Theorem. Finally, we use Proposition~\ref{main_thm} to obtain the bound~(\ref{r_rate3}) on the rates.\\
\underline{\emph{Step 1: Typical sets}}: We define the typical set
\begin{equation}\label{an_def}
A_n(\epsilon)= \left(A_{n,1}(\epsilon) \times  A_{n,2}(\epsilon)\right) \bigcap A_{n,3}(\epsilon)
\end{equation}
where
\[A_{n,1}(\epsilon) = \{x \in {\cal X}^{n}:  2^{-n(H(X)+\epsilon)} \leq p(x) \leq 2^{-n(H(X)-\epsilon)}\},\]
\[A_{n,2}(\epsilon) = \{y  \in {\cal Y}^{n} :  2^{-n(H(Y)+\epsilon)} \leq p(y) \leq 2^{-n(H(Y)-\epsilon)}\}\]
and
\begin{equation}
A_{n,3}(\epsilon) = \{(x,y) \in {\cal X}^{n} \times {\cal Y}^{n} :  2^{-n(H(X,Y)+\epsilon)} \leq p(x,y) \leq 2^{-n(H(X,Y)-\epsilon)}\} \nonumber
\end{equation}
with the notation that if~\(x=(x_1,\ldots,x_n),\) then~\(p(x) := \prod_{i=1}^{n}p(x_i).\)

\begin{figure}[tbp]
\centering
\includegraphics[width=3in, trim= 50 350 50 150, clip=true]{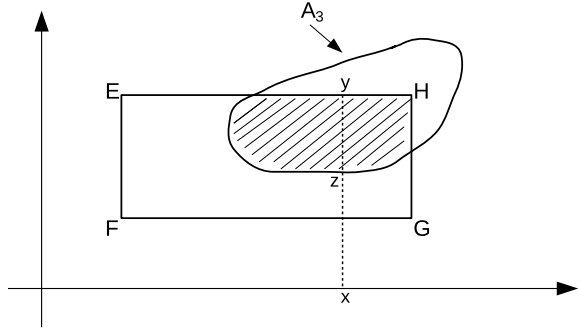}
\caption{The set~\(D_n(x,\epsilon)\) obtained from the sets~\(A_{n,i}(\epsilon), 1 \leq i \leq 3.\)}
\label{typ_fig}
\end{figure}

We estimate~\(\mathbb{P}(A_{n,1}(\epsilon))\) as follows. If~\((X_1,\ldots,X_n)\) is a random element of~\({\cal X}^{n}\) with~\(\{X_i\}\) i.i.d. and each having distribution~\(p(.),\) then the random variable~\(\log{p(X_i)}\) has mean~\(H(X)\) and so by Chebychev's inequality
\begin{eqnarray}
\mathbb{P}(A^c_{n,1}(\epsilon)) &=& \mathbb{P}\left( \left|\sum_{i=1}^{n} \log{p(X_i)} - nH(X) \right|\geq n H(X) \epsilon\right) \nonumber\\
&\leq& \frac{1}{n^2H^2(X) \epsilon^2} \mathbb{E}\left(\sum_{i=1}^{n} \log{p(X_i)} - nH(X) \right)^2 \nonumber\\
&=& \frac{1}{n H^2(X) \epsilon^2} \mathbb{E}\left(\log{p(X_1)}-H(X)\right)^2 \nonumber
\end{eqnarray}
which converges to zero as~\(n \rightarrow \infty.\) Analogous estimates hold for the sets~\(A_{n,2}(\epsilon)\) and~\(A_{n,3}(\epsilon)\)
and so
\begin{equation}\label{wln_est}
\mathbb{P}(A_n(\epsilon))  \geq 1-\epsilon^2
\end{equation} for all~\(n\) large.\\\\
\emph{\underline{Step 2: Determining the set~\({\cal B}_n\)}}: We now use the set~\(A_n(\epsilon)\) defined above to determine the set~\({\cal B}_n\) in the statement of the Theorem as follows. For~\(x \in A_{n,1}(\epsilon),\) let
\[D_n(x,\epsilon) := \{y \in A_{n,2}(\epsilon) : (x,y) \in A_n(\epsilon)\}.\]
In Figure~\ref{typ_fig}, we illustrate the sets~\(\{A_{n,i}(\epsilon)\}_{1 \leq i \leq 3}\) and the set~\(A_n(\epsilon).\)
The  rectangle~\(EFGH\) denotes~\(A_{n,1}(\epsilon) \times A_{n,2}(\epsilon)\) and the oval set~\(A_3\) represents~\(A_{n,3}(\epsilon).\)
The hatched region represents~\(A_n(\epsilon).\) The line~\(yz\) represents the set~\(D_n(x,\epsilon)\) for~\(x \in A_{n,1}(\epsilon)\) shown on the~\(X-\)axis.

From Figure~\ref{typ_fig} we see that
\begin{equation}\label{wln2}
\sum_{x \in A_{n,1}(\epsilon)} \left(\sum_{y \in D_n(x,\epsilon)} p(x,y)\right) = \sum_{(x,y) \in A_n(\epsilon)} p(x,y) \geq 1-\epsilon^2
\end{equation}
by~(\ref{wln_est}). Letting
\begin{equation}\label{an4_def}
A_{n,4}(\epsilon) := \left\{x \in A_{n,1}(\epsilon) : \sum_{y \in D_n(x,\epsilon)} p(y|x) \geq 1-\epsilon \right\},
\end{equation}
we split the summation in first term in~(\ref{wln2}) as~\(L_1+L_2\) where
\begin{equation}
L_1 = \sum_{x \in A_{n,4}(\epsilon)}\left(\sum_{y \in D_n(x,\epsilon)} p(y|x) \right) p(x) \leq \sum_{x \in A_{n,4}(\epsilon)} p(x)= \mathbb{P}\left(A_{n,4}(\epsilon)\right) \label{l1_est}
\end{equation}
and
\begin{eqnarray}
L_2 &=&  \sum_{x \in A_{n,1}(\epsilon) \setminus A_{n,4}(\epsilon)}\left(\sum_{y \in D_n(x,\epsilon)} p(y|x)\right)p(x) \nonumber\\
&\leq& (1-\epsilon) \sum_{x \in A_{n,1}(\epsilon) \setminus A_{n,4}(\epsilon)} p(x)\nonumber\\
&\leq& (1-\epsilon)\mathbb{P}\left(A^c_{n,4}(\epsilon)\right). \label{l2_est}
\end{eqnarray}

Substituting~(\ref{l2_est}) and~(\ref{l1_est}) into~(\ref{wln2}) we get
\[ 1-\epsilon \cdot \mathbb{P}\left(A^c_{n,4}(\epsilon)\right) \geq L_1 + L_2 \geq 1-\epsilon^2\] and so~\(\mathbb{P}\left(A^c_{n,4}(\epsilon)\right) \leq \epsilon.\) Because~\(A_{n,4}(\epsilon) \subseteq A_{n,1}(\epsilon),\) we therefore get
that~\[1-\epsilon \leq \mathbb{P}\left(A_{n,4}(\epsilon)\right) = \sum_{x \in A_{n,4}(\epsilon)} p(x)  \leq 2^{-n(H(X)-\epsilon)} \#A_{n,4}(\epsilon).\] Setting~\({\cal B}_n = A_{n,4}(\epsilon)\) we then get
\[\#{\cal B}_n \geq 2^{n(H(X)-\epsilon)} \cdot (1-\epsilon) \geq 2^{n(H(X)-2\epsilon)}\]
for all~\(n\) large.\\\\
\emph{\underline{Step 3: Using Proposition~\ref{main_thm}}}: For~\(\alpha \leq H(X),\) we let~\({\cal D}_n\) be any set of size~\(2^{n(\alpha-2\epsilon)}\) contained within~\({\cal B}_n.\) Let~\(G\) be the bipartite graph with vertex set~\({\cal X}_c \cup {\cal Y}_c\) where~\({\cal X}_c := {\cal D}_n,\;{\cal Y}_c := A_{n,2}(\epsilon) \)
and an edge is present between~\(x \in {\cal X}_c\) and~\(y \in {\cal Y}_c\) if and only if~\((x,y) \in A_n(\epsilon).\)
We now compute the sizes of the probable sets and the conflict sets in that order.

For each~\(x \in {\cal X}_c\) we have by definition~(\ref{an4_def}) of~\(A_{n,4}(\epsilon)\) that
\begin{equation}\label{dnr}
\sum_{y \in D_n(x,\epsilon)}p(y|x) \geq 1-\epsilon
\end{equation}
and so we set~\(D_n(x,\epsilon)\) to be the~\(\epsilon-\)probable set corresponding to~\(x \in {\cal D}_n.\)
To estimate the size of~\(D_n(x,\epsilon),\) we use the fact that~\((x,y) \in A_n(\epsilon)\) and so
\begin{equation}\label{pyx_est}
p(y|x) = \frac{p(x,y)}{p(x)} \geq \frac{2^{-n(H(X,Y)+\epsilon)}}{2^{-n(H(X)-\epsilon)}} = 2^{-n(H(Y|X)+2\epsilon)}.
\end{equation}
Thus~\[1 \geq \sum_{y \in D_n(x,\epsilon)}p(y|x) \geq  \#D_n(x,\epsilon) \cdot 2^{-n(H(Y|X)+2\epsilon)}\]
and consequently
\begin{equation}\label{dn_size}
\#D_n(x,\epsilon) \leq 2^{n(H(Y|X) + 2\epsilon)}.
\end{equation}

Finally, we estimate the size of the conflict set~\(C(y,\epsilon)\) for each~\(y \in {\cal Y}_c.\)
Again we use the fact that if~\((x,y)\) is an edge in~\(G\) then~\((x,y) \in A_n(\epsilon)\) and so
\begin{equation}\label{pyx_est2}
p(x|y) = \frac{p(x,y)}{p(y)} \geq \frac{2^{-n(H(X,Y)+\epsilon)}}{2^{-n(H(Y)-\epsilon)}} = 2^{-n(H(X|Y)+2\epsilon)}.
\end{equation}
Thus~\[ 1 \geq \sum_{x \in C(y,\epsilon)}p(x|y) \geq  \#C(y,\epsilon) \cdot 2^{-n(H(X|Y)+2\epsilon)}\]
and we get that~\(\#C(y,\epsilon) \leq 2^{n(H(X|Y) + 2\epsilon)}.\) Using this and~(\ref{dn_size}), we get that the conditions in Proposition~\ref{main_thm}
hold with~\[N_0 = 2^{n(\alpha -2\epsilon)}, d_L(\epsilon) = 2^{n(H(Y|X) + 2\epsilon)} \text{ and }d_R(\epsilon) = 2^{n(H(X|Y) + 2\epsilon)}.\] If~\(M = 2^{nR}\) with~\(R< \alpha -H(X|Y) - H(Y|X) -7\epsilon,\) then~(\ref{m_est}) holds
and so there exists a code containing~\(M = 2^{nR}\) words from~\({\cal D}_n\) giving an error probability
of at most~\(\epsilon\) with the conflict set decoder.\;\;\;\;\;\;\;\;\;\;\;\;\;\;\;\;\;\;\;\;\;\;\;\;
\;\;\;\;\;\;\;\;\;\;\;\;\;\;\;\;\;\;\;\;\;\;\;\;\;\;\;\;\;\;\;\;\;\;\;~\(\qed\)

\subsubsection*{Acknowledgements}
I thank Professors Rajesh Sundaresan, C. R. Subramanian and the referees for crucial comments that led to an improvement of the paper. I also thank IMSc for my fellowships.

\bibliographystyle{plain}

\end{document}